\documentstyle[11pt,aaspp4]{article}

\def\gtrapprox{\;\lower 0.5ex\hbox{$\buildrel >
    \over \sim\ $}}             
\def\lessapprox{\;\lower 0.5ex\hbox{$\buildrel < \over \sim\ $}}
\def\etal{et al.\ }

\def\mic{~$\mu$m\/}

\begin{document}

\title{ISOCAM 15\mic\ Search for Distant Infrared Galaxies Lensed
by Clusters}

\author{Richard Barvainis}
\affil{MIT Haystack Observatory, Westford, MA 01886 USA\footnote{
Current address: 4301 Columbia Pike \#610, Arlington, VA 22204}}
\centerline{Email: reb@patriot.net}

\author{Robert Antonucci}
\affil{UCSB, Physics Dept., Santa Barbara, CA 93106}
\centerline{Email: antonucci@physics.ucsb.edu}

\author{George Helou}
\affil{IPAC, 100-22, Pasadena, CA 91125}
\centerline{Email: gxh@ipac.caltech.edu}

\begin{abstract}
In a search for lensed infrared galaxies, ISOCAM images have been
obtained toward the rich clusters Abell 2218 and Abell 2219 at 15\mic .
Nine galaxies (four in Abell 2218 and five in Abell 2219) were
detected with flux levels in the range 530--1100 $\mu$Jy.  Three of
the galaxies detected in Abell 2218 have previously known redshifts;
of these one is a foreground galaxy and the other two are lensed
background galaxies at $z=0.474$ and $z=1.032$.  One of the
objects detected in the field of Abell 2219 is a faint, optically
red, extreme infrared-dominated galaxy with a probable redshift
of 1.048.

\end{abstract}

\keywords{Galaxies: Clusters} 

\section{Introduction}

  Clusters of galaxies can serve as windows on the distant universe by
bringing faint objects above detection threshholds via gravitational
lensing.  Many arcs representing magnified ordinary galaxies at moderate
to high redshifts are seen in rich galaxy clusters, and these have been
well-studied over the past several years.  Indeed some of the most
distant known galaxies have been discovered in this way at optical
wavelengths (see Franx et al 1997 for a lensed galaxy at $z=4.92$).
Clusters are known to be equally useful as cosmic magnifiers in other
wavebands as well.

  In the local universe the dominant populations of luminous objects
are infrared galaxies and quasars.  The ultraluminous infrared
galaxies (ULIRGs), with bolometric luminosities $\gtrsim 10^{12}
L_{\sun}$, appear to be powered by intense starbursts and obscured
quasars (e.g., Sanders and Mirabel 1996).  It remains an open
question how many such objects exist among galaxy populations
at high redshift, since they are obscured in the optical region
and generate most of their luminosity in the mid- to far-infrared
where sensitivities of space-borne telescopes such as IRAS have been
marginal for their detection above redshifts of a few tenths.  Indeed
only one such high-$z$ galaxy is known in the IRAS database, the infrared
galaxy/hidden quasar F10214+4724, at $z=2.3$ (Rowan-Robinson et
al 1991).  Two high-redshift IRAS quasars have also been identified:
the Cloverleaf at $z=2.6$, and APM 08279+5255 at $z=3.9$, both broad
absorption line quasars (Barvainis et al 1995; Irwin et al 1998).
Boosting by gravitational lensing was required for the detection
of F10214+4724  (Broadhurst and Leh\'ar 1995),
the Cloverleaf is a quad lens, and APM 08279+5255 appears to be a
close optical double and very likely lensed as well.  New lensed
iand unlensed infrared galaxies are currently being found via their
submillimeter emission by SCUBA on the JCMT
(Smail, Ivison, \& Blain 1997; Ivison et al 1998; Hughes et al 1998;
Barger et al 1998).

   ESA's Infrared Space Observatory (ISO) offered a new opportunity
to mount systematic searches for other high-$z$ infrared-dominated
objects.  The approach we adopted with ISO, which we report here,
was to use the mid-infrared camera ISOCAM to image at 15\mic\ two
very rich lensing clusters of galaxies, Abell 2218 and Abell 2219.
The strategy was to enhance detectability of distant
infrared galaxies over random fields by taking advantage of the
cluster lensing boost.  Other groups have carried out similar
programs: Altieri et al (1998a) for Abell 2218; Altieri et al (1998b)
and L\'emonon et al (1998) for Abell 2390; and Metcalfe et al (1998)
for Abell 370.

  The two target clusters were chosen for their richness and for
previous optical indications of lensing, along with purely
observational considerations such as visibility to ISO and high
ecliptic latitude.  Both are at moderate redshift: $z = 0.176$ for
Abell 2218, and $z = 0.225$ for Abell 2219.  In this experiment
several probable cluster galaxies were detected at 15\mic\ in
the two clusters.  Three background galaxies, dominated by their
mid-infrared emission, were also detected.  Two are previously
known lensed galaxies behind Abell 2218 at redshifts $z=0.474$
and $z=1.034$, and the third is a faint red object in the field of
Abell 2219 with a probable redshift of $z = 1.048$.

\section{Observations and Data Analysis}

Each cluster was observed for a total of 1.2 hours using the LW3 filter
of ISOCAM covering the range $12-18$\mic .  The 32$\times 32$ detector
array was configured with a plate scale of $6''$ per pixel.  Data
were taken in ``micro-scan" mode, using 3x3 rasters with $14''=2.33$
pixel steps to maximize the area covered and the different pixel
sampling of the same sky area.  Three separate such rasters, with
slightly shifted centers, were taken of each cluster, to achieve
yet more cross-sampling between camera pixels and sky pixels.
The diffraction-limited beamsize was $6.3''$ FWHM, but with a $6''$
pixel size and multiple sampling, we estimate that the point source
width should be $8-9''$.  There are no strong sources in the field
with which to measure an accurate point spread function, but the program
sources that were detected with reasonable SNR yield a FWHM of $9''$
when fitted with a circular Gaussian approximation to the PSF, 
or about the expected value.

For Abell 2218 the first raster was rendered unusable by detector instabilities.
Only the second and third rasters were summed, for a total integration
time of 0.8 hr.  For Abell 2219 all three rasters were acceptable.
Images were dark-subtracted, flat-fielded,
deglitched, corrected for transient response, and combined using
standard CIA routines at IPAC in Pasadena.

Source fluxes were derived by fitting with a circular two-dimensional
Gaussian function and zero level using AIPS, with fixed FWHM of $9''$
(see above).  The RMS noise levels, obtained by fitting with the
same Gaussian at random points in the fields, are $\approx 200$
$\mu$Jy for Abell 2218 and $\approx 110$ $\mu$Jy for Abell 2219.
These noise levels are well above the theoretical expectation, for
reasons that are not understood by us at present; the cause is not
a scaling error, since our source fluxes are consistent with those
of Alieri et al (1998a) (see \S3.1).

Absolute coordinates on the original ISO rasters have an uncertainty of
$\sim 6''$, and small shifts were required for both ISO images to
align the ISO sources with galaxies in optical frames.  No rotations
were required.   After the shifts all of the ISO sources had clear
optical identifications.

One of the detected sources in the Abell 2219 image (designated
A2219\#5, see below) is optically faint and very red, making it a
candidate for the sort of distant infrared-dominated galaxy we were
searching for.  A spectrum obtained for us by T.\ Broadhurst and
B.\ Frye using the Keck I telescope shows an emission
line  at $\lambda$7634\AA , to which we assign a probable
identification of [OII] $\lambda\lambda 3726,3729$\AA\ at $z =
1.048$.  This identification is supported by the continuum shape and the
lack of other strong lines in the $6300-9500$\AA\ passband observed
(see, e.g., starburst galaxy templates of Kinney et al 1996).
A detection of H$\alpha$
at 1.34~$\mu$m would confirm the redshift.  A possible alternative
identification would be H$\alpha$ at $z=0.163$, but we feel this
is less likely than [OII] at $z=1.048$ for the reasons given.

\section{Results and Discussion}

The 15\mic\ images are shown in Figures 1 and 2.  The coordinates shown
have been adjusted after comparison between the sources in the ISO image
and galaxies in I-band optical images, and
should be accurate to $1-2''$.  In Abell 2218 there are four sources with
detections for which we can be reasonably confident, and in Abell 2219 we
identify five sources.  Only detections that were evident on more
than one frame are considered bona fide sources here.  All of the
accepted sources are closely coincident with visible-light sources.
Contour overlays of the 15\mic\ images on $I$-band images from
the Palomar 5m telescope are shown in Figures 3 and 4 (5m images
courtesy I.\ Smail).  An overlay of the 15\mic\ image of Abell 2218 on
an HST $R$-band image is shown in Figure 5.  Object identifications,
with 15\mic\ and $I$-band fluxes, are given in Table 1.  All sources
are consistent with being point-like in the ISO images.

\subsection{Detected galaxies: General discussion}

Of the nine objects detected at 15\mic , three have redshifts
available in the literature.  These are A2218\#395, \#317, and \#289
(using the galaxy numbering system of Le Borgne, Pello, \& Sanahuja 1992;
Abell 2219 has no previous numbering system and the ISO sources are
numbered here according to RA order).
The latter two (see \S3.2) are background galaxies at $z = 0.474$ and
$z = 1.034$ respectively (Ebbels et al 1997), with \#289 clearly
being distorted by the cluster potential.  Abell 2218 has a mean redshift
of 0.176 (Le Borgne et al 1992).  A2218\#395, at $z=0.103$, is a
foreground galaxy and not part of the cluster.  As discussed above,
we assign a tentative redshift of $z=1.048$ for A2219\#5.

The other detected galaxies in Abell 2218 and Abell 2219 are of
unknown redshift and precise Hubble type, although the majority
appear to be spirals or irregulars.  None of those galaxies look
particularly distorted as if by lensing, and all except A2219\#5 have
optical colors typical of ordinary spiral or elliptical galaxies.
However, they are distinguished by their strong 15\mic\ emission,
which sets them apart from the hundreds of other galaxies in the
ISOCAM fields.

  The apparent mid-infrared luminosities (over the $\Delta \nu/\nu \sim
0.25$ of the filter passband) of the four galaxies with known
redshifts, A2218\#395, \#317, and \#289, and A2219\#5, are respectively
$L_{14\mu{\rm m}} = 3.9\times 10^8 L_{\sun}$,
$L_{10\mu{\rm m}} = 4.9\times 10^{9} L_{\sun}$, 
$L_{7\mu{\rm m}} = 3.0\times 10^{10} L_{\sun}$, 
and $L_{7\mu{\rm m}} = 1.9\times 10^{10} L_{\sun}$ (assuming $z=1.048$ 
for A2219\#5), 
where the subscripts on the luminosities
represent the rest wavelengths observed.  The frequency width of the
15\mic\ filter has been taken to be $5.0\times 10^{12}$ Hz, and the
cosmological parameters used are $H_0 = 65$ km s$^{-1}$ Mpc$^{-1}$
and $q_0 = 0.1$.  For A2218\#289 and \#317, and A2219\#5, the
calculated luminosities represent upper limits, since the fluxes are
likely to be magnified by lensing.  
The other galaxies, if assumed to be at the redshifts of their
clusters, range in luminosity between $5.8\times 10^8 L_{\sun}$
and $2.4\times 10^9 L_{\sun}$.  For comparison, of the 57 Virgo
cluster galaxies of all Hubble types detected by Boselli et al (1998)
at 15\mic , none have $L_{15\mu{\rm m}}$ above $10^9 L_{\sun}$,
and most lie in the range $10^6 - 10^8 L_{\sun}$.   
The galaxies detected here have much higher apparent luminosities
in the mid-infrared than average cluster galaxies.

Boselli et al (1998) find that for spirals and irregulars the
15\mic\ flux tends to be dominated by dust re-emission of primarily
UV stellar light, whereas for ellipticals it is the (very weak)
direct light of the Rayleigh-Jeans tail of the old population of
red stars.  However, for spirals the relation between star formation
and 15\mic\ flux is not simple.  It appears that galaxies with
moderate activity have the highest 15\mic\ to UV flux ratios whereas
for active star-forming galaxies this ratio is lower (Boselli et
al 1997).  The galaxies with clear spiral morphology in our sample
are A2218\#395 and \#317, and A2219\#1 and \#3.  A2218\#289 and
\#275 appear irregular.  The others, all in Abell 2219 (galaxies \#2,
\#4, and \#5), are difficult to classify because of lack of angular
resolution (all are smaller than $3''$).  However, the very small
number of detections of E/SO galaxies in Virgo relative to later
types by Boselli et al (1998) suggests that all or almost all of
the detections here are likely to be from spirals or irregulars.
This is supported by the large 15\mic\ to I-band flux ratios,
which  are typical of spirals rather than ellipticals (see Table 1).

Altieri et al (1998a) recently reported ISOCAM imaging of Abell 2218 at
5, 7, 10, and 15\mic .   Their field was roughly one-quarter the
size of ours, because their pixel field of view was  $3''$ compared
with our $6''$.  In the inner regions of the cluster they detected
the central cD galaxy and five others: \#395, \#317, \#323, \#373,
and \#275.  We detected \#395, \#317, \#275, and possibly \#373
(which we do not claim as a firm detection, although it does show
up with $\sim 400~\mu$Jy in two positive contours in Figures
4 \& 5).  A2218\#323 is weak in Figure 2 of Altieri et al, and
appears to be below our detection threshhold, as does the cD galaxy.
For \#395  and \#317 the fluxes given by Altieri et al (1998a)
are consistent with ours.  Our approximate flux of $\sim 400~\mu$Jy
for \#373 is also consistent with the value found by Altieri et al.

\subsection{Notes on individual galaxies}

  A2218\#289, at $z = 1.034$, consists of a complex of bright
knots and diffuse emission (see Figure 5).  Lensing distortion is
substantial at the northeastern end, where the galaxy is stretched across
the halo of cluster galaxy \#244; because the object appears so
luminous it is probably highly magnified, according to Kneib et al
(1996).

 A2218\#317 is a background spiral galaxy at $z = 0.474$.  It is
very probably magnified by the cluster, since its shape and color led to
a redshift prediction by Kneib et al (1996), using cluster inversion
techniques, of $0.2 < z_{\rm lensing} < 0.4$.  Given that  such redshift
predictions are statistical in nature, the spectroscopically measured
redshift of $z=0.474$ can be considered to be consistent with the
lensing hypothesis.

 A2219\#5 is optically the faintest object among those detected
at 15\mic , by more than an order of magnitude.  It is quite red,
with an optical color $B-I = 4.08$, and a 15\mic\ to $I$ ratio
$S_{15\mu {\rm m}}/S_I = 180$; the second largest such
ratio is 21 for A2218\#289.  A2219\#5 therefore appears to be
an unusual object, very red in the optical and apparently dominated
in luminosity by its mid- or far-infrared emission.  L\'emonon et
al (1998) and Altieri et al (1998b) have recently found similar
objects in ISOCAM images of Abell 2390.

Referrring to Figure 6, A2219\#5 is well-resolved in the optical
along the major axis and barely resolved along the minor axis.
A 2-dimensional gaussian fit gives a FWHM size of $2.0''\times 1.0''$
at position angle 103$\arcdeg$ (major axis).  The image seeing
is $\approx 0.7''$, so the deconvolved minor axis width would be
$\approx 0.7''$.  The direction to the center of the cluster from
the location of A2219\#5 lies at PA = 210$\arcdeg$ (see Figure 6),
so the major axis is within 17$\arcdeg$ of being perpendicular
to this direction, suggesting a lensed arclet.  

\subsection{IR/B Ratios}


Table 1 lists the 15\mic -to-I ratios for all detected galaxies.
The three galaxies at the highest measured  redshifts have the highest
ratios, as would be expected for a survey with fewer detections in
the infrared than the visible.  For A2218\#289 and A2219\#5 at
redshifts near 1, the observed 15\mic /I ratio maps closely to the
ratio of 7\mic /B in the rest frame.  For these bands, the observed
flux density ratios of 21 and 176 translate to a luminosity ratio
(in $\nu f_{\nu}$) of $\sim 1$ and $\sim 10$.

Helou et al (1999) show that for star forming galaxies the
ISO 7\mic\ band ($5-9$\mic ) is dominated by the Aromatic features,
and that the luminosity in this band accounts for $2-8\%$ of
the total infrared luminosity between 3 and 1000\mic .  The lower
luminosity fraction is characteristic of intense star bursts,
while the higher fractions occur in quiescent or mildly active star
forming galaxies.  At the extreme luminosity and excitation end
however, these fractions can drop even lower, as in Arp~220, which
emits only about 0.3\% of its luminosity in the 7\mic\ band (E. Sturm,
private communication; Genzel \etal 1998).  Using the higher 7\mic
-to-total-IR fraction of 8\% leads to IR/B ratios of 13 and 120
for A2218\#289 and A2219\#5, which are much higher than expected
in quiescent galaxies.  One is therefore led by self-consistency
arguments to using a smaller fraction applicable to active galaxies.
For a 7\mic -to-total-IR fraction of 2\%, one finds the total IR/B
ratios are $\sim 40$ and $\sim 400$ for A2218\#289 and A2219\#5 
(extrapolating the optical spectrum to obtain B$_{rest}$), and
the total observed IR luminosities about $1.5\times 10^{12}$
and $9.5\times 10^{11} L_{\sun}$.  While these are substantial
luminosities, they do have to be adjusted down for gravitational
lensing amplification, which is on the order of ten for A2218\#289
(Casoli et al 1996), and at most a few for A2219\#5 given that it is
lensed by the smooth cluster potential.  If A2219\#5 is at 
a redshift of 0.163 rather than 1.048 (see \S2), its luminosity
would be considerably lower of course but its IR/B ratio would
still be very large.

To further constrain the IR/B ratios we have analyzed the raw
IRAS survey data for the best estimate of upper limits in the
far-infrared, and find for each of A2218\#289 and A2219\#5 a
three-sigma upper limit of about 225 mJy at 100\mic , in quiet
sky with very little cirrus noise.  This implies that the 7\mic
-to-total-IR fraction must be greater than 1.6\% for A2218\#289,
and greater than 1\% for A2219\#5, or they would have been detected
by IRAS.  These limit fractions add confidence that the 2\% adopted
above is a reasonably good estimate, since it is constrained from
both above and below.  We therefore conclude that both of the $z=1$
objects are likely to be dusty galaxies similar to local ULIRGs,
with luminosities up to a few $\times 10^{11} L_{\sun}$.

While IR/B $\sim 40$ (A2218\#289) is not unusual for an active
galaxy, a ratio of 400 (A2219\#5) is extreme, in that it requires
a high column density of gas and dust with the dust surrounding the
active regions without significant leaks or clumping.  Differential
lensing amplification between wavelengths (Eisenhardt et al 1996)
is unlikely to be the cause of the extreme ratio since A2219\#5 is
lensed by the diffuse potential of Abell 2219, which is too smooth to
produce such an effect.  Similar high ratios have been reported by
Dey et al (1999) for the z=1.44 ERO (extremely red object) HR10 (IR/B
$\approx  300$), and by Yun and Scoville (1998) for IRAS F15307+3252
(IR/B $\approx  650$).

A2219\#5 has a number of other similarities to HR10 and other members
of the ERO class, although its colors are probably not  extreme
enough to warrant inclusion in that class (we estimate an R$-$K color
of 4.5, interpolating between our measured fluxes;
EROs have R$-$K$>6$).  The EROs are a population of infrared bright,
extremely red objects that have been discovered in near-infrared
imaging surveys.  Most EROs are sufficiently faint optically that
their redshifts and spectral properties are unknown.  The exception
is HR10, for which a redshift of 1.44 has been derived based on [OII]
and H$\alpha$ emission lines (Graham \& Dey 1996; Dey et al 1999).
This object is thought to be a high-redshift counterpart of the
dusty, ultraluminous infrared galaxies found in the local Universe
(Dey et al 1999).  The optical/IR spectral energy distributions
and luminosities of A2219\#5 (assuming $z=1.048$) and HR10 are
similar, both are extended at the $1-2''$ level, and both show an
emission line at long optical wavelengths, firmly identified as [OII]
$\lambda \lambda 3726,3729$\AA\ in the case of HR10 and tentatively
identified as the same line in A2219\#5.  If EROs are like A2219\#5
in their infrared properties, the mid-infrared ISO detection of
A2219\#5 indicates that the sharply rising near-infrared spectrum
of objects like HR10 continues out to at least 15\mic .  This is
consistent with the detection of HR10 at submillimeter wavelengths
by Dey et al (1999).

  We conclude that the galaxies in the cluster backgrounds are
luminous, high IR/B objects.   As is the case for most ULIRGs, the
primary driving power source, whether QSO or starburst, is unclear.

\section{Summary}

  Nine galaxies have been detected in ISOCAM 15\mic\
images of the lensing clusters Abell 2218 and Abell 2219.  At least
three of these are luminous, high IR/B lensed background galaxies at
redshifts of $0.5-1$, and one, behind Abell 2219, is an optically faint
object heavily dominated by its mid-infrared emission.  Judging from
its high infrared luminosity, its high ratio of infrared to optical
emission, and its red optical colors, this object appears to be
active, with the source or sources of activity (starburst and/or AGN)
heavily obscured by dust.

\acknowledgments 
We thank Ian Smail for kindly providing optical images and photometry
of the clusters, Jean-Paul Kneib for the HST image, and 
Tom Broadhurst and Brenda Frye for obtaining the Keck spectrum of
A2219\#5.

\newpage
\vglue 1.0truecm
\centerline {\bf FIGURE CAPTIONS}

\figcaption{ISOCAM 15\mic\ image of Abell 2218.  Pixel size is $2''$, and
gray scale range is $-40$ to $+40$ $\mu$Jy/pixel.  Note coordinate rotation
of $40\arcdeg$ clockwise (see tick marks).  Horizontal striping near the bottom
of the image is an artifact. \label{fig1}}

\figcaption{Same as Figure 1, for Abell 2219.  Coordinate rotation
$44\arcdeg$ clockwise. \label{fig2}}

\figcaption{Contour overlay of ISOCAM image on I-band image from
the Palomar 5-m telescope for Abell 2218.  Dashed contours are negative.
Contour levels:
($-3,-2,-1, 1, 2, 3, 4$)$\times 9$ $\mu$Jy per $2''$ pixel in original image.
\label{fig3}}

\figcaption{Same as Figure 3, for Abell 2219.   
Contour levels:
($-3,-2,-1, 1, 2, 3, 4, 5, 6$)$\times 7$ $\mu$Jy per $2''$ pixel in
original image.  \label{fig4}}

\figcaption{Contour overlay of ISOCAM image on R-band (F702W) HST image
for Abell 2218.  Contours are the same as in Figure 3.  For A2218\#289 the
centroids of the optical and infrared images differ by $< 2''$;  the
northwest extending lower contours for this object are caused by the
image artifact noted in Figure 1.  HST image courtesy J.-P. Kneib
(see Kneib et al 1996)\label{fig5}}

\figcaption{ Close-up of A2219\#5 from the 5-m I-band image
(Figure 4).  Direction to cluster center is almost perpendicular
to galaxy major axis, suggesting that this object
may be a lensed arclet.  \label{fig6}}

\begin{deluxetable}{lccccccccccc}
\tablenum{1}
\tablewidth{0pt}
\footnotesize
\tablecaption{Galaxies Detected at 15\mic }
\tablehead{
\colhead{Cluster} &\colhead{ID\tablenotemark{a}} 
&\colhead{RA} &\colhead{Dec} &\colhead{$z$\tablenotemark{b}} 
&\colhead{$S_{15\mu {\rm m}}$} 
&\colhead{$S_I$\tablenotemark{c}} 
&\colhead{$S_V$} 
&\colhead{$S_B$} 
&\colhead{$S_U$} 
&\colhead{$B-I$} 
&\colhead{$S_{15\mu{\rm m}}/S_I$} 
\nl
\colhead{~} &\colhead{~} 
&\colhead{(2000)} &\colhead{(2000)} &\colhead{~} 
&\colhead{($\mu$Jy)} &\colhead{($\mu$Jy)} 
&\colhead{($\mu$Jy)} &\colhead{($\mu$Jy)} 
&\colhead{($\mu$Jy)} 
&\colhead{(mag)} 
}
\startdata
A2218 &\#395& 16 35 48.8 & 66 13 02 & 0.103  &$1100\pm
200$&223&99&73&27&1.90&5\nl
A2218 &\#317& 16 35 53.3 & 66 12 58 & 0.474  &$670\pm
200$&45&11&6&2&2.96&15\nl
A2218 &\#289& 16 35 54.8 & 66 11 52 & 1.034  &$850\pm
200$&41&14&8&4&2.52&21\nl
A2218 &\#275& 16 35 55.4 & 66 14 15 &  ...   &$570\pm
200$&66&23&16&10&2.20&9\nl
A2219 & \#1 & 16 40 13.0 & 46 43 07 &  ...   &$890\pm
110$&125&...&37&17&2.02&7\nl
A2219 & \#2 & 16 40 13.9 & 46 41 51 &  ...   &$920\pm
110$&504&...&83&22&2.65&2\nl
A2219 & \#3 & 16 40 21.5 & 46 40 49 &  ...   &$1420\pm
110$&201&...&40&14&2.45&7\nl
A2219 & \#4 & 16 40 22.5 & 46 43 12 &  ...   &$1100\pm
110$&86&...&12&5&2.79&13\nl
A2219 & \#5 & 16 40 23.0 & 46 44 03 &1.048&$530\pm
110$&3&...&0.13&$<0.1$&4.08&176\nl
\tablenotetext{a} {Using numbering system of Le Borgne, Pello, \& 
Sanahuja (1992) for Abell 2218, or RA order for Abell 2219.}  
\tablenotetext{b} {Redshifts for objects in Abell 2218 from Ebbels et al
(1997); redshift for A2219\#5 is tentative, based on an emission line
identified as [OII] $\lambda\lambda 3726,3729$\AA\ (see \S2 of this
paper).}
\tablenotetext{c} {Integrated flux densities of galaxies.  Estimated
errors $\approx 5\%$ except for A2219\#5, where the errors are 8\% at I
and 15\% at B.  For A2218\#289, the photometry covers the main galaxy, and 
does not include the thin ``tail'' extending to the northeast (see Figure
5).  The $I$-band photmetry is Kron-Cousins $I$, with effective
wavelength 8000\AA , and an assumed zero point of 2380 Janskys.
The other bands are standard Johnson $U, B, V$.
Photometry courtesy Ian Smail.}

\enddata 
\end{deluxetable} 

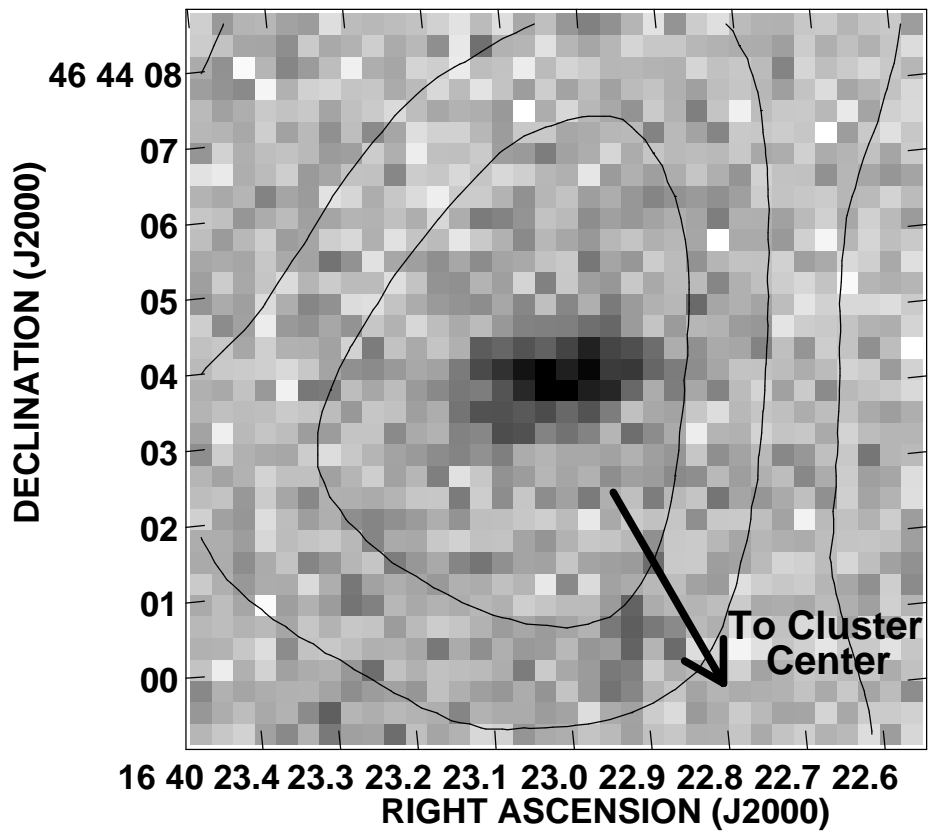
\end{document}